\documentclass[twocolumn]{revtex4}
\usepackage{amsmath}
\usepackage{amssymb}
\usepackage{rsfs}
\usepackage{color}
\usepackage{ulem}

\newcommand{\ra}{\rangle}
\newcommand{\la}{\langle}

\begin{document}

\title{Lorentzian holographic gravity and the time--energy uncertainty principle}

\author{Eiji Konishi}
\email{konishi.eiji.27c@kyoto-u.jp}
\address{Graduate School of Human and Environmental Studies, Kyoto University, Kyoto 606-8501, Japan}

\date{\today}

\begin{abstract}
In this article, we present a heuristic derivation of the on-shell equation of the Lorentzian classicalized holographic tensor network in the presence of a non-zero mass in the bulk spacetime.
This derivation of the on-shell equation is based on two physical assumptions.
First, the Lorentzian bulk theory is in the ground state.
Second, the law of Lorentzian holographic gravity is identified with the time--energy uncertainty principle.
The arguments in this derivation could lead to a novel picture of Lorentzian gravity as a quantum mechanical time uncertainty based on the holographic principle and classicalization.
\end{abstract}

\maketitle

\section{Introduction}


The holographic principle in the theory of gravity \cite{Hol1,Hol2,Hol3} equates the degrees of freedom of the bulk spacetime with the information stored in the quantum field defined on the boundary spacetime.
The $d+2$-dimensional anti-de Sitter spacetime/$d+1$-dimensional conformal field theory (AdS$_{d+2}$/CFT$_{d+1}$) correspondence is the known spacetime example of the holographic principle \cite{AdSCFT1,AdSCFT2}.


In a tensor network model of the bulk AdS$_3$ spacetime (i.e., the case where $d=1$) \cite{Swingle,MIH,Bao,Review1,Review2}, the classicalization of the quantum pure state $|\psi\ra$ (the ground state) of the holographic tensor network (HTN) of a strongly coupled boundary CFT$_2$ has been proposed and investigated by the present author \cite{EPL1,EPL2}.
Here, {\it the tensor network model} refers to the multi-scale entanglement renormalization ansatz (MERA) of the boundary quantum pure state $|\psi\ra$ \cite{Vidal1,Vidal2}.
The MERA is a scale-invariant tensor network consisting of the semi-infinitely alternate combinations of the disentanglement layer (disentanglers) and the real-space renormalization layer (isometries) in the radial direction.
This is called a {\it holographic} tensor network because it corroborates the Ryu--Takayanagi holographic formula for the subregion entanglement entropy of the boundary CFT$_2$ \cite{RT,Book} when the radial direction in the MERA is identified with the holographic direction.

We classicalize the quantum pure state $|\psi\ra$ of the boundary CFT$_2$ by introducing the superselection rule operator $\sigma_3$ (the Pauli third matrix) in the qubits' eigenbasis of the boundary CFT$_2$ into the set of the observables \cite{JHAP1}.
Here, in quantum mechanics, the {\it superselection rule} selects the observables that commute with a given superselection rule operator \cite{WWW,Jauch}.
Applying this rule, the quantum coherence (i.e., the off-diagonal part of the density matrix) in the eigenbasis of the superselection rule operator is exactly and completely lost.
Namely, after the classicalization, the density matrix $|\psi\ra\la \psi|$ becomes equivalent to an exactly diagonal matrix (i.e., a classical mixed state) in the qubits' eigenbasis with respect to the restricted set of the observables.


In the Euclidean and Lorentzian regimes of the bulk spacetime, the bulk action of the classicalized holographic tensor network (cHTN) of the ground state $|\psi\ra$ was proposed in Ref. \cite{EPL2} as
\begin{equation}
I_{\rm bulk}[|\psi\ra]=-\hbar H[|\psi\ra]\;,\label{eq:action}
\end{equation}
where $H[|\psi\ra]$ is the Shannon entropy (in nats) of the diagonal density matrix $|\psi\ra\la \psi|$ with respect to the restricted set of the observables and quantifies the amount of the information lost in the ground state of the boundary CFT$_2$ by its classicalization.


On the basis of Eq. (\ref{eq:action}), the present author previously investigated the holographic gravity induced by a non-zero mass in the bulk spacetime in the Euclidean regime \cite{JHAP2}.
There, the bulk action of the mass is identified with information readings from the cHTN, and this identification leads to the Euclidean holographic gravity induced by the mass as the Unruh effect \cite{Review2,JHAP2,Sewell,Review3}.
However, in the Lorentzian regime, because the sign of the bulk action of a non-zero mass is negative, such informatical identification does not exist.

To resolve this tension, in the presence of a non-zero mass $M$ at the top tensor of the HTN in the ground state, the on-shell equation of the full bulk action in the Lorentzian regime was derived in Ref. \cite{JHAP3}.
The result is
\begin{equation}
-\sigma \hbar \theta =Mc^2\;,\label{eq:JHAP4}
\end{equation}
where $\sigma$ is the entropy production (in nats per site of the HTN) accompanying the classicalization, and
\begin{equation}
\theta=\frac{1}{d^2A}\frac{d}{d\tau} d^2A\label{eq:theta}
\end{equation}
is the real-proper-time expansion of the infinitesimal bulk-space area $d^2A$, surrounding the site occupied by the mass (i.e., the top tensor of the HTN) in the bulk space, for the real proper time $\tau$.

In Ref. \cite{JHAP3}, this on-shell equation (\ref{eq:JHAP4}) is identified as the equation for the Lorentzian holographic gravity induced by the mass by invoking the Unruh effect \cite{Review2,Sewell,Review3}.


However, the physical and intuitive meaning of this on-shell equation (\ref{eq:JHAP4}) itself has not yet been revealed.
The purpose of this article is to present a heuristic derivation of this on-shell equation (\ref{eq:JHAP4}) with clear physical meaning.


The rest of the article is organized as follows.
In the next section, we present a heuristic derivation of Eq. (\ref{eq:JHAP4}) based on the two physical assumptions of the Lorentzian bulk theory.
In the final section, we conclude the article.

\section{A heuristic derivation of Eq. (\ref{eq:JHAP4})}


Our starting points are the following two physical assumptions of the Lorentzian bulk theory:
\begin{enumerate}
\item[(I)] The Lorentzian bulk theory is in the ground state.
That is, the full Lorentzian bulk action is equivalent to Eq. (\ref{eq:action}) on the shell of the cHTN \cite{JHAP3}.
In particular, the rest energy of a mass is regarded as a quantum mechanical energy uncertainty of the cHTN in the ground state.

\item[(II)] The law of Lorentzian holographic gravity is identified with the time--energy uncertainty principle.
\end{enumerate}
These assumptions constitute our novel picture of Lorentzian holographic gravity.


Here, we give an overview of the concept of the time--energy uncertainty principle in non-relativistic quantum mechanics \cite{Heisenberg,LP,MT,Messiah,Jammer}.
In its primitive form, this principle asserts the relation
\begin{equation}
\Delta E \cdot \Delta T \gtrsim \hbar
\end{equation}
for the energy uncertainty $\Delta E$ of a quantum mechanical system and the time interval $\Delta T$ during which the energy uncertainty $\Delta E$ is maintained.


The treatment of the time uncertainty $\Delta T$ in this form was advanced by Mandelstam and Tamm to a more rigorous form in Ref. \cite{MT}.
The Mandelstam--Tamm time--energy uncertainty relation in non-relativistic quantum mechanics is given by
\begin{equation}
\Delta E \cdot \Delta R\gtrsim \frac{\hbar}{2} \left|\frac{d\la \widehat{R}\ra}{d t}\right|\label{eq:MT}
\end{equation}
for a dynamical variable operator $\widehat{R}$ of the system and a coordinate time $t$.
Importantly, the Mandelstam--Tamm time uncertainty 
\begin{equation}
\tau_R=\Delta R\cdot \left|\frac{d\la \widehat{R}\ra}{d t}\right|^{-1}
\end{equation}
is the time required for the quantum mechanical expectation value $\la \widehat{R}\ra$ of the dynamical variable $R$ to show its minimal distinct change \cite{MT,Messiah,Jammer}.
Namely, during the time interval $\tau_R$, the uncertainty $\Delta R$ of $R$ is maintained.
This form of the principle (\ref{eq:MT}) can be derived from the fundamental postulates in non-relativistic quantum mechanics \cite{Messiah}.


Now, we apply the time--energy uncertainty principle to our case of the cHTN in the Lorentzian regime.
In this case, from assumption (I), we regard the rest energy of the mass as the energy uncertainty.
That is,
\begin{equation}
\Delta E=Mc^2\;.
\end{equation}
In addition to this, we consider
\begin{equation}
\Delta T \simeq \tau_R\label{eq:DeltaT}
\end{equation}
and
\begin{equation}
R=d^2A\;,\ \ \Delta R=d^2A
\end{equation}
for the dynamical variable $R$ of the cHTN and its uncertainty $\Delta R$.
Namely, the Mandelstam--Tamm time uncertainty $\tau_R$ is the characteristic proper time for $d^2A$ (i.e., the cHTN) to show its minimal distinct change.\footnote{Here, we assume that $\theta d\tau$ is always negative.
In Ref. \cite{JHAP3}, this assumption is derived from the full Lorentzian bulk action by its minimization.}
Specifically, $\tau_R$ is the proper time required to shift the labels of the discrete inverse renormalization group steps in the cHTN by minus one \cite{JHAP3}.


Then, in the heuristic form, our time--energy uncertainty relation is
\begin{equation}
Mc^2 \cdot d^2A \gtrsim \hbar \left|\frac{d}{d\tau}d^2 A\right|\;.
\end{equation}
By using Eq. (\ref{eq:theta}), this relation can be rewritten as
\begin{equation}
Mc^2 \gtrsim -\hbar \theta\label{eq:TEUP}
\end{equation}
when $d^2A$ is temporally always decreasing (see the footnote).
The equality in Eq. (\ref{eq:TEUP}) is in the same form as our desired on-shell equation (\ref{eq:JHAP4}).

\section{Conclusion}


In this article, we presented a heuristic derivation of the on-shell equation of the Lorentzian cHTN in the presence of a non-zero mass in the bulk spacetime.
In Ref. \cite{JHAP3}, this on-shell equation is identified as the equation for the Lorentzian holographic gravity induced by the mass by invoking the Unruh effect \cite{Review2,Sewell,Review3}.


Our heuristic derivation of the on-shell equation is based on two physical assumptions.
First, the Lorentzian bulk theory is in the ground state.
Second, the law of Lorentzian holographic gravity is identified with the time--energy uncertainty principle.

In this derivation, the rest energy of the mass in the Lorentzian regime is identified with the quantum mechanical energy uncertainty of the cHTN in the ground state, whereas the action of the mass in the Euclidean regime is identified with information readings from the cHTN \cite{JHAP2}.
This difference between the two regimes stems from and is consistent with the different physical meanings of the action of the cHTN in these regimes.
Namely, in the Euclidean regime, the meaning of the Dirac constant $\hbar$ in the action of the cHTN is one classicalized spin degree of freedom.
In other words, this is the {\it holographic principle}.
On the other hand, in the Lorentzian regime, the meaning of the Dirac constant $\hbar$ in the action of the cHTN is the lower bound in the quantum mechanical uncertainty principle relations.
In other words, this is the {\it bulk quantum mechanics}.
In particular, the latter meaning of $\hbar$ requires the existence of a quantum measuring system in the bulk spacetime in the Lorentzian regime \cite{JHAP4}.

To conclude this article, we note that the present arguments could lead to a novel picture of Lorentzian gravity as a quantum mechanical time uncertainty based on the holographic principle and classicalization.

\end{document}